\documentclass[11pt]{article}
\usepackage{latexsym}

\newcommand{\beq}{\begin{equation}}
\newcommand{\eeq}{\end{equation}}
\newcommand{\beqs}{\begin{eqnarray}}
\newcommand{\eeqs}{\end{eqnarray}}
\newcommand{\tr}{\mathrm{tr\,}}
\newcommand{\WW}{{\cal W}}
\newcommand{\refe}[1]{(\ref{#1})}

\newcommand{\weff}{W_{\mathrm{eff}}}
\newcommand{\wtree}{W_{\mathrm{tree}}}
\newcommand{\wpert}{W_{\mathrm{pert}}}

\begin{document}
\begin{titlepage}
\vskip 2.5cm
\begin{center}
{\LARGE \bf Equivalence of effective superpotentials}
\vspace{2.71cm}

{\Large
Riccardo Argurio}
\vskip 0.5cm
{\large \it Physique Th\'eorique et Math\'ematique\\
and\\ 
International Solvay Institutes\\
\smallskip
Universit\'e Libre de Bruxelles, C.P. 231, 1050 Bruxelles, Belgium}
\vskip 0.5cm
{\tt email: Riccardo.Argurio@ulb.ac.be}
\end{center}
\vspace{2cm}
\begin{abstract}
We show that the low-energy effective superpotential of an ${\cal N}=1$
$U(N)$ gauge theory with matter in the adjoint and arbitrary
even tree-level superpotential has, in the classically unbroken case, 
the same functional form
as the effective superpotential of a $U(N)$ gauge theory with matter
in the fundamental and the same tree-level interactions,
up to some rescalings of the couplings. We also argue that the same kind
of reasoning can be applied to other cases as well.
\end{abstract}

\end{titlepage}

Recent research on the exact vacuum structure of ${\cal N}=1$
supersymmetric gauge theories has highlighted the central role 
played by the effective glueball superpotential $\weff(S)$, in particular
due to the systematic techniques developed in \cite{dv,dglvz,cdsw}
in order to compute it. For a review and a list of references, see \cite{afh}.

The object of this note is to show that seemingly different theories
can have an effective glueball superpotential $\weff$ with exactly the
same functional form.
We will discuss in some detail the case of the ${\cal N}=1$ $U(N)$
gauge theory with a fundamental-antifundamental pair $Q, \tilde Q$,
which shares an equivalent effective superpotential with the same gauge theory
with matter $\Phi$ in the adjoint and an even tree level superpotential 
$\wtree$, in the unbroken, maximally confining $U(N)$ vacuum. 
The proof of the equivalence uses the
generalized Konishi anomaly, more precisely its reduction to the
``ordinary'' Konishi anomaly in the case of interest.

A similar, but technically more involved, treatment could in principle
be given for other cases with matter in other
representations, for instance in the antisymmetric representation. In fact, 
the present investigation was spurred by the conjectured equivalence of
the effective superpotentials in this latter case in \cite{ags}, conjecture
based on the orientifold parent-daughter relation between the theory 
with the adjoint and the one with the antisymmetric. For $U(3)$, it becomes
a relation between theories with adjoint and with fundamental matter. 

We now discuss the determination of $\weff$ in the simpler case
of the fundamental matter. We find a generalization of the superpotential 
obtained in \cite{acfh},
using the method outlined in \cite{binor}, that is using the Konishi
anomaly equation.

Having one flavour, and calling the meson operator $X\equiv Q \tilde Q$,
the general $\wtree$ has a simple form:
\beq
\wtree = m X + {1\over 2} \lambda X^2 + \dots  
= \sum_k {1\over k} \lambda_k X^k.
\eeq
Such a superpotential can of course be the result of integrating
out other fields, e.g. in the adjoint, with Yukawa-like couplings to
the fundamental ones, as in the example discussed in \cite{acfh}.

The Konishi anomaly, basically the anomaly for the $U(1)_Q$
rotations, yields the following relations for the expectation values
in a SUSY vacuum:
\beq 
 \sum_k  \lambda_k X^k = S, \label{konfun}
\eeq
with:
\beq
S=-{1\over 32 \pi^2}\tr \WW^\alpha
\WW_\alpha = {1\over 16 \pi^2}\tr \lambda^\alpha \lambda_\alpha + \dots
\eeq
The relation \refe{konfun} is then solved to obtain $X(S,\lambda_k)$.
The reason why we can do this so easily is that, $X$ being a chiral
operator, its powers have the factorization property, i.e. in a SUSY vacuum:
\beq
\langle X^k \rangle =\langle X \rangle^k. \label{factor}
\eeq
Since there is no ambiguity, we drop the explicit reference to the VEV.

Having $X(S,\lambda_k)$, 
we can compute the effective superpotential by integrating
the equations:
\beq
{\partial \weff \over \partial \lambda_k}=  {1\over k}  X^k.
\eeq
Note that a coupling independent part of $\weff$ is left undetermined
(the VY piece), so we are determining this way what is called $\wpert$.

As an example, we can consider the case $\lambda_1=m$, $\lambda_2=\lambda$
and $\lambda_k=0$ for $k\geq 3$. Here solving for $X$ is simple enough,
and one can also integrate to find a closed expression for $\wpert$.
We have:
\beq
X={m\over 2\lambda}\left[ -1 + \sqrt{1+ 4 {\lambda S\over m^2}}\right],
\eeq
and:
\beq
\wpert= -{m^2\over 4\lambda}\left(1-\sqrt{1+ 4 {\lambda S\over m^2}}\right)
+S \log {1\over 2}\left(1+\sqrt{1+ 4 {\lambda S\over m^2}}\right)
+S\log m. \label{fundweff}
\eeq
In order to obtain the full $\weff$, one needs to add a VY piece \cite{vy}
reflecting the effective gauge dynamics at scales where the $SU(N)$
confines and the matter field is integrated out. Note that
in order to do this one has to carefully extract from $\wpert$ any
remaining $S$-dependence.

We now consider the case of the theory with adjoint matter, and an
even tree level superpotential:
\beq
\wtree= \sum_{k=2}^{n+1}{1\over k} g_{k} \tr \Phi^{k}, \qquad g_{2l+1}=0.
\eeq
The effective superpotential is also determined by the VEVs of
gauge invariant operators:
\beq
{\partial \weff \over \partial g_{2k}} = {1\over 2k} \langle  \tr \Phi^{2k}
\rangle.
\eeq
In this case however it is more involved to determine the VEVs, since in
general a relation like \refe{factor} is not available. Hence the 
Konishi anomaly related to simple $U(1)_\Phi$ rotations is not enough
to determine all the VEVs needed to extract $\weff$. We need to
generalize the Konishi anomaly relations, as shown in \cite{cdsw}.

Here we will remark that in an unbroken $U(N)$ vacuum, there are actually
factorization-like relations $\langle  \tr \Phi^{2k}\rangle \propto
\langle  \tr \Phi^2\rangle^k$. We proceed to show this purely in 
${\cal N}=1$ language.

We thus introduce the objects which
allow to write an infinite series of Konishi anomaly relations
in closed form \cite{cdsw}, the resolvents:
\beq
R(z) = -{1\over 32 \pi^2}\left\langle\tr {\WW^\alpha\WW_\alpha \over
z-\Phi }\right\rangle, \qquad T(z) =  \left\langle\tr {1  \over
z-\Phi }\right\rangle.
\eeq
In terms of the above, the anomaly relations read:
\beq
R(z)^2= \wtree'(z)R(z)+{1\over 4}f(z), \qquad 2R(z)T(z)=\wtree'(z)T(z) 
+{1\over 4}c(z), \label{genkon}
\eeq
where $\wtree'(z)=\sum_{k=1}^{n}g_{k+1}z^k$, and $f(z)$ and $c(z)$ are
polynomials of degree $n-1$.

The solution for $R(z)$ is simple enough:
\beq
2R(z) = \wtree'(z) - \sqrt{\wtree'(z)^2 + f(z)}.
\eeq
The basic feature of the above solution is the number of cuts on the 
complex plane it has. This allows to translate, in general, the data within
the polynomial $f(z)$ into the gluino bilinear VEVs for every
unbroken low-energy gauge group:
\beq
S_i={1\over 2\pi i}\oint_{C_i} dz\; R(z),
\eeq 
where $C_i$ circles the $i$-th cut. 

If we are now in the case that all the eigenvalues of $\Phi$ distribute
around one classical value (say, $\Phi=0$), the gauge group $U(N)$ is
unbroken, and we expect at the effective level only one glueball
effective superfield $S$. This corresponds to having only one cut
in the solution for $R(z)$. In other words, all the zeros of 
$\wtree'(z)^2 + f(z)$ must come in pairs except two.
We can thus write:
\beq 
\wtree'(z)^2 + f(z) = g_{n+1}^2(z^2-a_1^2)^2 \dots (z^2-a_{n-1\over 2}^2)^2
(z^2-b^2).
\eeq
Since $\wtree(z)$ is even, the problem is invariant
under reflection in $z$. We have thus taken an even $f(z)$ and 
imposed that for any zero there is a mirror zero under reflection.

We do not need to consider further the resolvent $R(z)$, since
what we need in order 
to compute $\weff$ are the first few terms in the expansion
of $T(z)$. From the anomaly equations \refe{genkon}, 
we see that $T(z)$ is given by:
\beq
T(z)=-{1\over 4} {c(z) \over  \wtree'(z)-2R(z)} = -{1\over 4} {c(z) \over 
\sqrt{\wtree'(z)^2 + f(z)}}.
\eeq
We recall now the conditions on $T(z)$, basically fixing the respective size of
the low energy gauge groups $U(N_i)$:
\beq
N_i = {1\over 2\pi i}\oint_{C_i} dz\;T(z).
\eeq
In the unbroken case, this means that the overall contribution
(the residue at infinity) comes entirely from the unique cut, 
while all residues at the simple poles must vanish.
This condition implies that $c(z)$ is such that it cancels all the simple
poles, and enforces that $T(z) \sim {N\over z}$ when $z \rightarrow \infty$.
Thus, the conclusion is that in the unbroken case $T(z)$ takes
a very simple form, depending on a single parameter $b^2$:
\beq 
T(z) = {N \over \sqrt{z^2-b^2}}.
\eeq
We have then a simple expression for the VEVs $\langle \tr \Phi^{2l}\rangle$,
all in terms of the single parameter $b^2$:
\beq
T(z)= \sum_{k\geq 0} {1\over z^{k+1}}\langle \tr \Phi^k\rangle
= {N\over z}\left(1-{b^2\over z^2}\right)^{-{1\over 2}}= 
{N\over z}\sum_{k\geq 0}{(2k)!\over 4^k (k!)^2} {b^{2k}\over z^{2k}}.
\label{expan}
\eeq
This implies:
\beq
\langle {1\over N}\tr \Phi^{2k}\rangle={(2k)!\over 4^k (k!)^2} b^{2k}
= {(2k)!\over 2^k (k!)^2} \langle{1\over N} \tr \Phi^2\rangle^k.
\label{relation}
\eeq
These relations have already been obtained through factorization of
the Seiberg-Witten curve \cite{ds}, see for instance \cite{ferrari}. Note that
these are {\em not} chiral ring relations, and that they are not satisfied
by giving a classical VEV to $\Phi$. Indeed, \refe{expan} also
implies $\langle \tr \Phi^{2l+1}\rangle =0$. 
In the following, 
it is useful to define $u_k=\langle {1\over N}\tr \Phi^{k}\rangle$.
For instance, we have the relation:
\beq
u_4= {3\over 2} u_2^2 .
\eeq
We can now use the relations \refe{relation} in order to solve for 
$u_2(S,g_k)$, using only the equation for the $U(1)_\Phi$ chiral anomaly:
\beq
\sum_k g_k \langle \tr \Phi^k\rangle = 2N S,
\eeq
which can also be written as:
\beq
\sum_k g_k u_k=2S. \label{adjkon}
\eeq
Inserting \refe{relation}, we obtain:
\beq
\sum_k g_{2k} {(2k)!\over 2^k (k!)^2} u_2^k = 2S.
\eeq
This means that the functional form of $u_2(S,g_{2k})$ is given by 
the functional form of $X(S,\lambda_k)$, by\footnote{Since $S$ and
$g_k$ will anyway appear in dimensionless combinations in $\weff$,
some numerical factors can actually be shifted from one variable
to another, consistently with \refe{adjkon}, 
without any change in the final result.}:
\beq
u_2(S,g_{2k}) = X(2S, {(2k)!\over 2^k (k!)^2}g_{2k}).
\eeq
Using the above expressions to obtain $\weff$ for the theory with
the adjoint, it can be expressed in terms of $\weff$ of the theory
with the fundamental as:
\beq
\weff^\mathrm{(Adj)}(S,g_{2k})={N\over 2} \weff^\mathrm{(Fund)}
(2S, {(2k)!\over 2^k (k!)^2}g_{2k}).
\eeq
As an example, consider the theory with:
\beq
\wtree={1\over 2}g_2 \tr \Phi^2 + {1\over 4}g_4 \tr \Phi^4.
\eeq
We have that:
\beq
u_2=X(2S,g_2,{3\over 2}g_4)=
{g_2\over 3 g_4}\left[ -1 + \sqrt{1+ 12 {g_4 S\over g_2^2}}\right].
\eeq
and:
\beq
\wpert= -N{g_2^2\over 12g_4}\left(1-\sqrt{1+ 12 {g_4 S\over g_2^2}}\right)
+NS \log {1\over 2}\left(1+\sqrt{1+ 12 {g_4 S\over g_2^2}}\right)
+NS\log g_2. \label{adjweff}
\eeq
This same superpotential can be obtained through a matrix model computation,
using for instance the results of \cite{bipz}.

A similar reasoning should be used to compare also with 
the effective superpotential for other theories, namely the one
with antisymmetric matter considered in \cite{ags}.
In that case, one should further be able to
analyze the multi-cut case (that is, 
when the gauge group is broken to two or more factors 
at low energies).

One can also use the techniques applied here to the adjoint theory in order
to compute $\weff$ in the more general case
of a not necessarily even $\wtree$. In the one-cut case, $T(z)$ will
depend on two parameters, the two edges of the cut. We will thus be
able to express all the gauge invariants $\langle \tr \Phi^k\rangle$ 
in terms of, say,
the first two. A relation among the latter two can be found using the
(traced) classical equations of motion. Eventually, the ordinary Konishi
anomaly will give us an equation for the remaining invariant in terms
of the couplings and $S$. One can check that it is possible to rederive
in this way the effective superpotential of the theory with a cubic
interaction (and hence the free energy of the related matrix model,
as given in \cite{bipz}).

A more formal consequence of the equivalence of the effective superpotentials
\refe{fundweff} and \refe{adjweff}, is also the functional equivalence
of the free energies of the related (one-cut) matrix models. The adjoint case
is simply related to the free energy of the matrix model with 
quartic potential computed in \cite{bipz}. The fundamental case on the other
hand can be related to a matrix model with boundaries \cite{acfh}, which
in turn can be translated to a matrix model with a logarithmic potential.
It would be nice to understand
this correspondence at the matrix model level.

\subsection*{Acknowledgments}
I would like to thank G.~Bonelli, G.~Ferretti and especially A.~Armoni
for very useful discussions and feedback.
This work is supported in part by the ``Actions de Recherche
Concert{\'e}es" of the ``Direction de la Recherche Scientifique -
Communaut{\'e} Fran{\c c}aise de Belgique", by a ``P\^ole
d'Attraction Interuniversitaire" (Belgium), by IISN-Belgium
(convention 4.4505.86)  and by the European Commission RTN programme
HPRN-CT-00131. The author is a Postdoctoral Researcher of
the Fonds National de la Recherche Scientifique (Belgium).

\end{document}